\documentclass[fleqn,usenatbib]{mnras}

\usepackage{newtxtext,newtxmath}

\usepackage[T1]{fontenc}
\usepackage{ae,aecompl}

\usepackage{graphicx}	
\usepackage{amsmath}	
\usepackage{amssymb}	
\usepackage{multirow}
\usepackage{longtable}

\newcommand{\Teff}{{T_\mathrm{eff}}}
\newcommand{\Tldr}{{T_\mathrm{LDR}}}
\newcommand{\Tkov}{{T_\mathrm{KOV}}}
\newcommand{\logg}{{\log g}}
\newcommand{\FeH}{\mathrm{[Fe/H]}}
\newcommand{\Pe}{P_\mathrm{e}}
\newcommand{\XL}{\mathrm{X_{low}}}
\newcommand{\XH}{\mathrm{X_{high}}}
\newcommand{\logLDR}{\log{\mathrm{LDR}}}
\mathchardef\mhyphen="2D

\title[The gravity effect on LDRs]{The effect of surface gravity on line-depth ratios in the wavelength range  0.97--1.32\,$\mu \mathrm{m}$}

\author[M. Jian et al.]{
	Mingjie Jian,$^{1}$\thanks{E-mail: mingjie@astron.s.u-tokyo.ac.jp}
    Daisuke Taniguchi,$^{1}$
	Noriyuki Matsunaga,$^{1,2}$
    Naoto Kobayashi,$^{2,3,4}$
    \newauthor
	Yuji Ikeda,$^{2,5}$
    Chikako Yasui,$^{2,6}$ 
    Sohei Kondo,$^{3,2}$ 
    Hiroaki Sameshima,$^{4,2}$
    \newauthor
	Satoshi Hamano,$^{2,6}$ 
    Kei Fukue,$^{2}$ 
    Akira Arai,$^{2}$ 
    Shogo Otsubo$^{2}$ and
    Hideyo Kawakita$^{2,7}$
\\
	$^{1}$Department of Astronomy, School of Science, The University of Tokyo, 7-3-1 Hongo, Bunkyo-ku, Tokyo 113-0033, Japan\\
	$^{2}$Laboratory of Infrared High-resolution spectroscopy (LiH), Koyama Astronomical Observatory, Kyoto Sangyo University, \\Motoyama,
	Kamigamo, Kita-ku, Kyoto 603-8555, Japan \\
    $^{3}$Kiso Observatory, Institute of Astronomy, School of Science, The University of Tokyo, 10762-30 Mitake, Kiso-machi, \\
    Kiso-gun, Nagano 397-0101, Japan\\
    $^{4}$Institute of Astronomy, School of Science, The University of Tokyo, 2-21-1 Osawa, Mitaka, Tokyo 181-0015, Japan\\
    $^{5}$Photocoding, 460-102 Iwakura-Nakamachi, Sakyo-ku, Kyoto 606-0025, Japan\\
    $^{6}$National Astronomical Observatory of Japan, 2-21-1 Osawa, Mitaka, Tokyo 181-8588, Japan\\
    $^{7}$Department of Astrophysics and Atmospheric Sciences, Faculty of Science, Kyoto Sangyo University, Motoyama, \\
    Kamigamo, Kita-ku, Kyoto 603-8555, Japan
}

\date{Accepted XXX. Received YYY; in original form ZZZ}

\pubyear{2020}

\begin{document}
\label{firstpage}
\pagerange{\pageref{firstpage}--\pageref{lastpage}}
\maketitle

\begin{abstract}

A line-depth ratio (LDR) of two spectral lines with different excitation potentials is expected to be correlated with the effective temperature ($\Teff$). 
It is possible to determine $\Teff$ of a star with a precision of tens of Kelvin if dozens or hundreds of tight LDR--$\Teff$ relations can be used.
Most of the previous studies on the LDR method were limited to optical wavelengths, but Taniguchi and collaborators reported 81 LDR relations in the $YJ$ band, 0.97--1.32\,$\mu\mathrm{m}$, in 2018.
However, with their sample of only 10 giants, it was impossible to account for the effects of surface gravity and metallicity on the LDRs well.
Here we investigate the gravity effect based on $YJ$-band spectra of 63 stars including dwarfs, giants, and supergiants observed with the WINERED spectrograph.
We found that some LDR--$\Teff$ relations show clear offsets between the sequence of dwarfs and those of giants/supergiants.
The difference between the ionization potentials of the elements considered in each line pair and the corresponding difference in the depths can, at least partly, explain the dependency of the LDR on the surface gravity.
In order to expand the stellar parameter ranges that the LDR method can cover with high precision, we obtained new sets of LDR--$\Teff$ relations for solar-metal G0--K4 dwarfs and F7--K5 supergiants, respectively.
The typical precision that can be achieved with our relations is 10--30\,K for both dwarfs and supergiants. 

\end{abstract}




\section{Introduction}
\label{sec:intro}

Determining stellar parameters, especially the effective temperature ($\Teff$), is a fundamental part of the analysis of stellar spectra.
There are several methods of determining $\Teff$.
For example, often used in recent large projects is searching for the synthetic spectrum with the optimized stellar parameters, including $\Teff$, that matches an observed spectrum best (e.g., with the ASPCAP, the pipeline used by the Apache Point Observatory Galactic Evolution Experiment, or APOGEE, survey; \citealt{garcia_perez_aspcap:_2016, APOGEE}).
Fitting synthetic spectra to observed ones is usually performed with one or more wide spectral ranges that include features sensitive to different stellar parameters such as abundances of various elements.
Therefore, many parameters need to be determined simultaneously, which increases the computational load and the chance and degree of degeneracy between various parameters.
Another approach is to use empirical relations between $\Teff$ and observational indices such as photometric colours \citep[e.g.,][]{Huang2015} and the H$\alpha$ index \citep[e.g.,][]{Joner2015}.
Measuring such indices is usually simple and does not require a large amount of computational resource.

Here we focus on line-depth ratios (LDRs) and LDR--$\Teff$ relations. 
Spectral lines with different excitation potentials (EPs) have different sensitivity to $\Teff$, which provides the possibility to use them for determining $\Teff$.
Having a large number of well-calibrated LDR--$\Teff$ relations allows us to achieve high precision of the final $\Teff$.
One of the advantages of the LDR method is that the relations are empirically calibrated and are not directly affected by uncertainties in stellar models and those in the list of absorption lines.
It is still possible that the calibration data include systematic errors, but the relations can be easily improved once better (more precise and robust) effective temperatures of the calibrating stars become available.
Following the pioneering studies of \citet{Gray1989} and \citet{Gray1991}, more than a hundred LDR relations were obtained for stars of different luminosity classes separately: dwarfs (\citealt{Kovtyukh2003, Kovtyukh2004}), giants (\citealt{Strassmeier2000, Gray2001, Kovtyukh2006a, Kovtyukh2006b}), and supergiants (\citealt{Kovtyukh1998, Kovtyukh2000, Kovtyukh2007}).
The uncertainties of the final $\Teff$ in the previous studies range from 5 to 30\,K at best cases.

Most previous works on the LDR method, like those mentioned above, have been done with optical spectra mostly shorter than 0.8\,$\mu\mathrm{m}$.
\citet{Sasselov1990} extended the application to the infrared range by investigating a few lines of C\,$\textsc{i}$ and Si\,$\textsc{i}$ in the $Y$ band (around 1.1\,$\mu\mathrm{m}$).
Recently, about a dozen of line pairs in the $H$-band ($1.51$--$1.70\,\mu\mathrm{m}$) and 81 line pairs in the $YJ$-band ($0.97$--$1.32\,\mu\mathrm{m}$) were found to give tight LDR--$\Teff$ relations as reported by \cite{Fukue}, \citet{Jian2019a}, and \citet[][hereafter T18]{Taniguchi}.

The metallicity and surface gravity of a star can also affect the line depths and LDRs, and thus the effects of these parameters need to be understood and taken into account.
The metallicity effect on LDR relations was first detected by \citet{Taylor1994}, and \citet{Gray1994} suggested that it is caused by the saturation of absorption lines.
Using only weak lines to avoid the saturation is expected to be useful in reducing the metallicity effect \citep{Gray1994, Kovtyukh2000}.
Recently, \citet{Jian2019a} clearly detected the metallicity effect on the $H$-band LDRs using the spectra of red giants observed in APOGEE.
They found that the high-EP line, at least, of every line pair investigated is saturated in a wide metallicity range ($\FeH > -1$\,dex); therefore, the metallicity effect cannot be eliminated by the approach mentioned above at least for their relations.
\citet{Jian2019a} thus introduced metallicity- and abundance ratio-dependent terms to their LDR relations.
On the other hand, the gravity effect was discussed, e.g., in \citet{Catalano2002} and \citet{biazzo_effective_2007}.
They obtained linear relations between the residuals with respect to fiducial LDR--$\Teff$ relations and the gravity indices ($\logg$ and absolute magnitude offset from the zero-age main sequence).
Then, their relations give gravity-corrected LDRs that can be used to estimate $\Teff$ of stars in a broad range of $\logg$.

In this paper, we first compare the LDR relations of dwarfs, giants, and supergiants for the line pairs found by T18 to investigate the gravity effect.
The $YJ$ band spectra of 63 stars were obtained with WINERED (Warm INfrared Echelle spectrograph to Realize Extreme Dispersion and sensitivity; \citealt{WINERED}), the same instrument used in T18.
We then present two sets of new LDR--$\Teff$ relations in the $YJ$-band, one for dwarfs and the other for supergiants.
Our sample consists of stars with metallicity within 0.2\,dex around the solar, and thus the metallicity effect is expected to be small if any.

The python package we used for measuring line depths and LDRs and for deriving $\Teff$ is provided as Supporting Information and also on the web\footnote{ \url{https://github.com/MingjieJian/ir_ldr}}.

\section{Data and analysis}
\label{sec:data}

\subsection{Targets and observations}
\label{sec:target}

We use the $YJ$-band spectra of 20 dwarfs, 25 giants, and 18 supergiants taken with WINERED.
They were observed with the WIDE-mode giving the resolution of around 28000.
The spectra between 0.91 and 1.35\,$\mu \mathrm{m}$ are covered with 20 echelle orders (from 42nd to 61st).
The observations were carried out with the 1.3\,m Araki Telescope at Koyama Observatory, Kyoto Sangyo University in Japan from July 2015 to May 2016.
A part of the spectra of giants and supergiants were used in \citet{Matsunaga2019} to identify absorption lines of neutron-capture elements.

\begin{table*}
    \centering
    \caption{Stellar parameters of the dwarfs in our sample together with the mean S/N of the objects (Obj) and the telluric standards (Tel) described in Section~\ref{sec:spec-redu} and the observation dates. The first 10 lines are presented, and the full table is available as Supporting Information.}
\begin{tabular}{ccccccc}
\hline
   Object &       $\Teff$ &       $\FeH$&     $\logg$ & \multicolumn{2}{c}{S/N} & Obs. date \\
          &         (K)   &        (dex) &      (dex)  & Obj & Tel & \\     
\hline
 HD219134 &   $4900 \pm 7.9^{[2]}$ &   $0.10^{[2]}$ &   $4.20^{[2]}$ &     681 &     532 &  2015.10.25 \\
  HD10476 &   $5242 \pm 3.2^{[2]}$ &   $0.00^{[2]}$ &   $4.30^{[2]}$ &     314 &     513 &  2015.10.31 \\
 HD101501 &   $5558 \pm 6.1^{[2]}$ &   $0.02^{[2]}$ &   $4.50^{[2]}$ &     494 &     309 &  2016.01.26 \\
 HD114710 &   $5954 \pm 6.8^{[2]}$ &   $0.12^{[2]}$ &   $4.30^{[2]}$ &     556 &     309 &  2016.01.26 \\
 HD115383 &   $6012 \pm 9.3^{[2]}$ &   $0.16^{[2]}$ &   $4.30^{[2]}$ &     541 &     309 &  2016.01.26 \\
 HD117176 &   $5611 \pm 4.7^{[1]}$ &  $-0.10^{[3]}$ &   $3.86^{[3]}$ &     488 &     309 &  2016.01.26 \\
 HD137107 &   $6037 \pm 6.9^{[2]}$ &   $0.05^{[2]}$ &   $4.30^{[2]}$ &     522 &     309 &  2016.01.26 \\
 HD143761 &  $5865 \pm 11.1^{[2]}$ &  $-0.06^{[2]}$ &   $4.30^{[2]}$ &     474 &     309 &  2016.01.26 \\
 HD102870 &   $6055 \pm 6.8^{[2]}$ &   $0.18^{[2]}$ &   $4.00^{[2]}$ &     355 &     578 &  2016.01.27 \\
  HD72905 &   $5884 \pm 6.8^{[2]}$ &  $-0.02^{[2]}$ &   $4.40^{[2]}$ &     481 &     524 &  2016.01.27 \\
 HD122064 &   $4937 \pm 8.1^{[2]}$ &   $0.12^{[2]}$ &   $4.50^{[2]}$ &     402 &     452 &  2016.02.03 \\
\hline
\end{tabular}

    References: 1.~\citet{Kovtyukh2003}; 2.~\citet{Kovtyukh2004}; 3.~\citet{Lee2011}.

\label{tab:paras_dwarf}
\end{table*}

\begin{table*}
    \centering
    \caption{Stellar parameters of the giants in our sample together with the mean S/N of the objects (Obj) and the telluric standards (Tel) described in Section~\ref{sec:spec-redu} and the observation dates. We adopted $\Teff$  from \citet{Kovtyukh2006b} and other parameters as indicated in the table. The first 10 lines are presented, and the full table is available as Supporting Information.}
\begin{tabular}{ccccccc}
\hline
   Object &       $\Teff$ &       $\FeH$&     $\logg$ & \multicolumn{2}{c}{S/N} & Obs. date \\
          &         (K)   &        (dex) &      (dex)  & Obj & Tel & \\        
\hline
  HD11559 &        $4977 \pm 7.4$ &   $0.16^{[4]}$ &   $3.23^{[4]}$ &     481 &     598 &  2015.10.23 \\
  HD27348 &        $5003 \pm 6.1$ &   $0.05^{[2]}$ &   $2.75^{[2]}$ &     555 &     305 &  2015.10.25 \\
  HD27697 &        $4975 \pm 7.6$ &   $0.12^{[2]}$ &   $2.64^{[2]}$ &     516 &     305 &  2015.10.25 \\
  HD28292 &        $4453 \pm 9.0$ &  $-0.08^{[4]}$ &   $2.81^{[4]}$ &     523 &     305 &  2015.10.25 \\
  HD28305 &        $4925 \pm 8.7$ &   $0.20^{[3]}$ &   $2.72^{[3]}$ &     414 &     305 &  2015.10.25 \\
 HD198149 &        $4858 \pm 8.1$ &  $-0.19^{[3]}$ &   $3.29^{[3]}$ &     275 &     304 &  2015.10.26 \\
  HD25604 &        $4764 \pm 7.4$ &   $0.07^{[4]}$ &   $2.73^{[4]}$ &     335 &     403 &  2015.10.28 \\
  HD27371 &        $4960 \pm 8.1$ &   $0.15^{[3]}$ &   $2.76^{[3]}$ &     445 &     403 &  2015.10.28 \\
  HD48433 &        $4471 \pm 7.8$ &  $-0.16^{[3]}$ &   $2.10^{[3]}$ &     302 &     403 &  2015.10.28 \\
  HD76813 &        $5060 \pm 5.5$ &  $-0.06^{[3]}$ &   $2.63^{[3]}$ &     529 &     403 &  2015.10.28 \\
\hline
\end{tabular}

    References:  1.~\citet{Park2013}; 2.~\citet{Liu2014}; 3.~\citet{Prugniel2011}; 4.~\citet{da-Silva2015}.
    
\label{tab:paras_giant}
\end{table*}

\begin{table*}
    \centering
    \caption{Stellar parameters of the supergiants together with the mean S/N of the objects (Obj) and the telluric standards (Tel) described in Section~\ref{sec:spec-redu} and the observation dates. We adopted $\Teff$ from \citet{Kovtyukh2007} and other parameters as indicated in the table. The first 10 lines are presented, and the full table is available as Supporting Information.}
\begin{tabular}{ccccccc}
\hline
   Object &       $\Teff$ &       $\FeH$&     $\logg$ & \multicolumn{2}{c}{S/N} & Obs. date \\
          &         (K)   &        (dex) &      (dex)  & Obj & Tel & \\   
\hline
 HD194093 &  $6202 \pm 11.5$ &   $0.05^{[1]}$ &   $1.35^{[1]}$ &     348 &     304 &  2015.10.26 \\
 HD204867 &   $5466 \pm 7.4$ &   $0.03^{[1]}$ &   $1.54^{[1]}$ &     429 &     885 &  2015.10.26 \\
  HD26630 &   $5337 \pm 6.6$ &   $0.09^{[1]}$ &   $1.74^{[1]}$ &     387 &     403 &  2015.10.28 \\
  HD37536 &  $3918 \pm 76.2$ &   $0.13^{[3]}$ &   $0.52^{[3]}$ &     449 &     403 &  2015.10.28 \\
  HD48329 &  $4510 \pm 10.8$ &   $0.12^{[3]}$ &   $0.92^{[3]}$ &     460 &     403 &  2015.10.28 \\
  HD52005 &  $3954 \pm 37.6$ &   $0.13^{[3]}$ &   $0.78^{[3]}$ &     408 &     403 &  2015.10.28 \\
 HD206778 &  $4108 \pm 30.1$ &   $0.11^{[3]}$ &   $0.84^{[3]}$ &     415 &     513 &  2015.10.31 \\
  HD31910 &  $5441 \pm 10.1$ &  $-0.01^{[1]}$ &   $1.57^{[1]}$ &     310 &     513 &  2015.10.31 \\
   HD3421 &  $5302 \pm 10.1$ &  $-0.20^{[2]}$ &   $1.88^{[2]}$ &     209 &     513 &  2015.10.31 \\
   HD9900 &  $4529 \pm 15.6$ &   $0.19^{[1]}$ &   $1.35^{[1]}$ &     272 &     513 &  2015.10.31 \\
\hline
\end{tabular}

    References:  1.~\citet{Luck2014}; 2.~\citet{Liu2014}; 3.~\citet{Lee2011}.
    
\label{tab:paras_spg}
\end{table*}

\begin{figure}
	\centering
	\includegraphics[width=0.95\columnwidth]{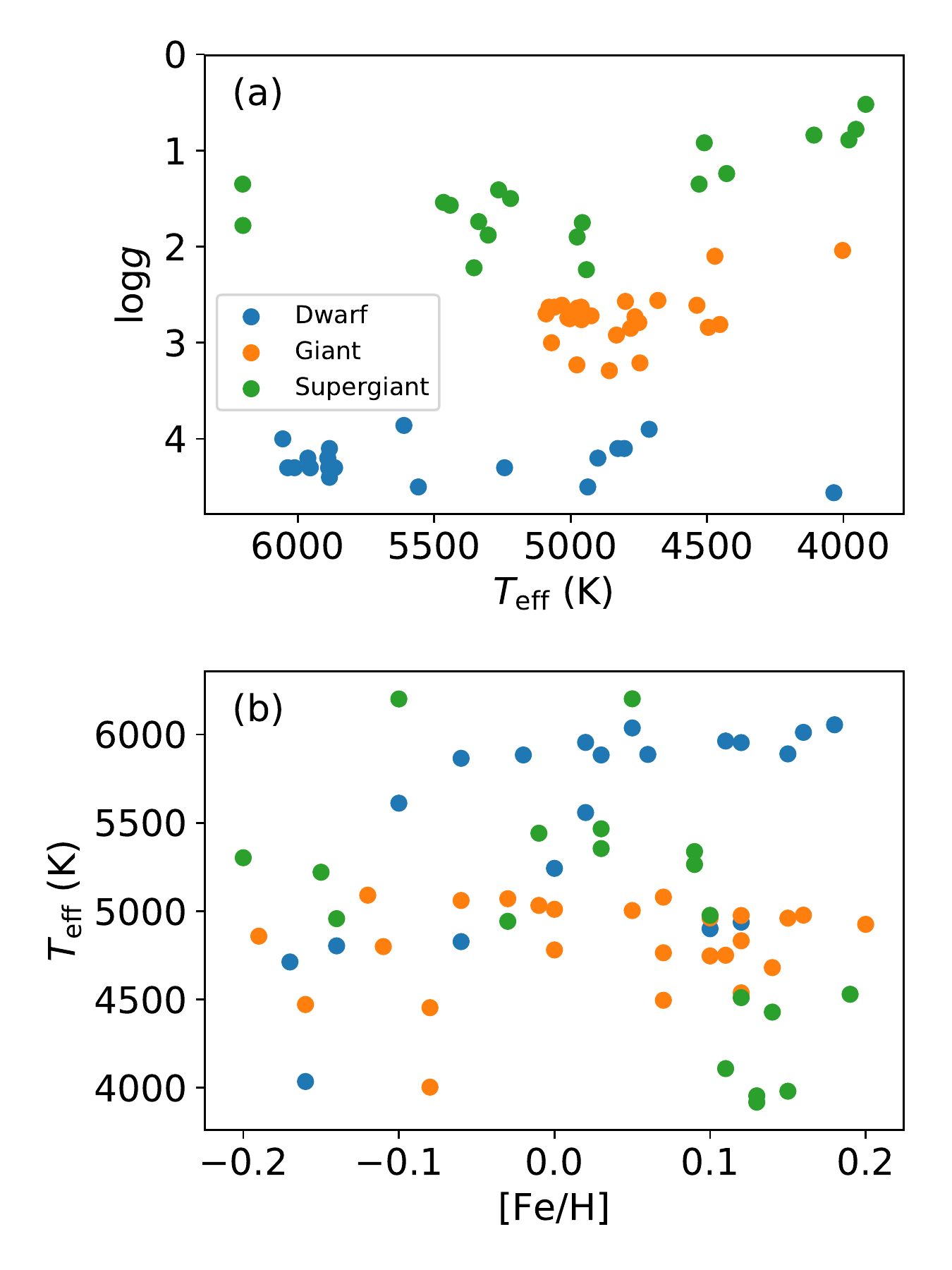}
	\caption{Parameters of our target sample on (a) the $\logg$--$\Teff$ plane (Kiel diagram), and (b) the $\Teff$--[Fe/H] plane.}
	\label{fig:paras}
\end{figure}

The target stars were selected from the catalogues of \citet{Kovtyukh2003, Kovtyukh2004, Kovtyukh2006b} and \citet{Kovtyukh2007}.
Kovtyukh and collaborators determined $\Teff$ of a large number of stars based on $\sim$\,100 optical LDR--$\Teff$ relations for dwarfs, giants, and supergiants with the precision around 10\,K.
The catalogues of \citet{Kovtyukh2003, Kovtyukh2006b} and \citet{Kovtyukh2007} do not provide $\FeH$ and $\logg$, and we adopted the metallicities and $\logg$ of the objects from other studies as listed in Table~\ref{tab:paras_dwarf}--\ref{tab:paras_spg}.
Fig.~\ref{fig:paras} plots the stellar parameters of our sample.
The dwarfs occupy a narrow gravity range on the Kiel diagram at around $\logg \approx 4.2$\,dex, and their $\Teff$ range from $4000$ to $6000$\,K.
The giants are distributed along the red giant branch with a clump at around $\Teff = 5000$\,K.
Some of the stars at the clump may actually be red clump stars at the central helium burning phase, but they are all treated as the same group of giants in the following analysis.
The supergiants are spread between $4000$ and $6100$\,K  in $\Teff$ and between $0.5$ and $2.0$\,dex in $\logg$.
To minimize the metallicity effect, we limited the metallicity of our sample to around the solar ($-0.2 < \FeH < 0.2$) as illustrated in Fig.~\ref{fig:paras}(b).

\subsection{Spectral reduction}
\label{sec:spec-redu}

\begin{figure*}
	\centering
	\includegraphics[width=2\columnwidth]{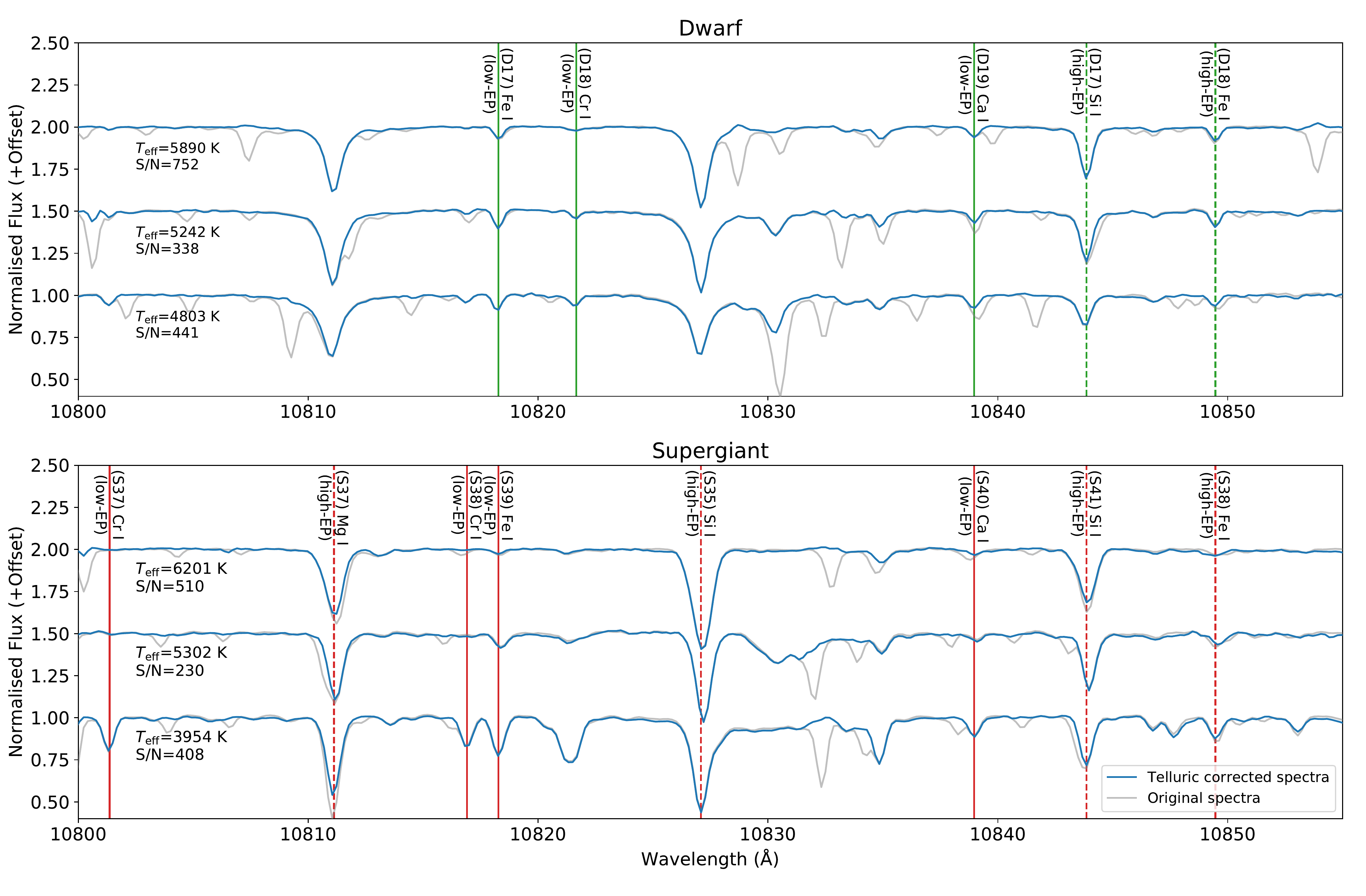}
	\caption{Examples of dwarf and supergiant spectra.  Sections of the telluric-corrected and original spectra from the order 52nd are coloured in blue and grey. The low-EP and high-EP lines in the dwarf and supergiant line pairs (obtained in Section~\ref{sec:new-relations}) are marked with solid and dashed vertical lines, respectively.}
	\label{fig:spectra-example}
\end{figure*}

The raw spectral data of the observed stars were first reduced by the pipeline developed by the WINERED team (Hamano et al. in preparation).
Every star was observed with more than one exposures, and the multiple spectra of each order were combined by the pipeline with their relative wavelength offsets\footnote{The wavelength shifts are, at least partly, due to the instability of the instrument caused by varying ambient temperature, and we made an instrumental upgrade to minimize such shifts in late 2016.}, if any, corrected.
The pipeline outputs the normalized spectra along with supplementary information such as the signal-to-noise ratio (S/N) estimates.
The mean S/N of the orders 43rd--48th and 52nd--57th for each star and the counterpart for the telluric standard star are listed in Table~\ref{tab:paras_dwarf}--\ref{tab:paras_spg}. 
The telluric correction was then performed using the spectra of telluric standard stars with their intrinsic spectral lines removed according to the method of \citet{Sameshima_telluric}.
The orders 42nd, 49th--51st, and 58th--61st are seriously affected by the telluric absorption.
They were not used by T18 for establishing the LDR relations, and thus we also excluded these orders.
In contrast, the telluric absorption in the orders of 53rd and 54th is weak and no telluric correction was applied to them.
Since the continuum level of a telluric-corrected spectrum is often biased from the unity, the continuum normalisation using the $\textsc{IRAF~continuum}$ task was applied to each order of the spectra after the telluric correction.
Finally, we corrected wavelength shifts of all the individual orders of the telluric corrected spectra separately to place stellar lines at the rest wavelengths in the standard air.
Examples of dwarf and supergiant spectra from the order 52nd, before and after telluric correction, are presented in Fig.~\ref{fig:spectra-example}.   
Throughout this paper, we use air wavelengths rather than vacuum wavelengths.

\subsection{Line depth measurements}
\label{sec:LDR-measure}

The line depths were measured as follows (see \citealt{Jian2019a} for more details).
For each line, 4 or 5 pixels around the centre were fitted by a parabola, and we took the distance between the minimum of the parabola and the continuum (1 in the normalized flux scale) as the depth.
We excluded the line depths if the fitting was bad or if the position of the minimum was offset from the line centre by more than 0.2\,\AA.
We note that the measurements tend to fail when the lines get too shallow, often, in stars with higher temperature.
The error of a line depth was derived considering the S/N of each order (not the mean S/N given in Table~\ref{tab:paras_dwarf}--\ref{tab:paras_spg}).
For the orders of 53rd and 54th, the S/N of the object spectrum give the error, $e_\mathrm{total} = e_\mathrm{Obj} = (\mathrm{S/N_{Obj}})^{-1}$.
In case of the other orders for which the telluric correction was made, the error caused by the telluric spectrum, $e_\mathrm{Tel} = (\mathrm{S/N_{Tel}})^{-1}$ is added in the quadrature to give $e_\mathrm{total} = (e_\mathrm{Obj}^2 + e_\mathrm{Tel}^2)^{1/2}$.
The relative errors (error divided by depth) were around 10 per cent or less for most of the depths measured.
We then calculated the LDR and its error for each line pair.

\section{Gravity effect on LDR relations}
\label{sec:gravity-effect}

In this section, we compare the depths and LDRs of dwarfs, giants, and supergiants to study the gravity effect on the T18 LDR--$\Teff$ relations. 
The line pair IDs used in this section are adopted from the tables 4 and 5 in T18 and prefixed by `T'.
The trends of the LDRs for giants seen in our data set are consistent with the LDR--$\Teff$ relations in T18.
However, we found that the distributions on the LDR--$\Teff$ diagram show systematic offsets between the three groups of luminosity class for some line pairs while the offsets are not significant for others.

In the following discussion, we first limit ourselves to the LDRs whose lines are not significantly blended with other lines and discuss the gravity effect on such LDRs.
Whilst the LDRs affected by some line blends still show tight relations with $\Teff$, it is important to understand how the gravity affects the lines of LDR pairs themselves.
In order to evaluate the degrees of line blends of the T18 line pairs, we considered a set of synthetic spectra with $(\Teff, \logg) = (4500, 1.0), (4500, 2.5), (5000, 1.5), (5000, 2.5)$, and $(5000, 4.0)$.
The spectral synthesis was performed using MOOG (a code that performs LTE line analysis and spectrum synthesis, \citealt{MOOG}) with the Vienna Atomic Line Database (VALD3 line list; \citealt{VALD3}) and the ATLAS9 stellar models from \citet{meszaros_new_2012}.
We considered the ratio $d_\mathrm{syn}^- / d_\mathrm{syn}$ where $d_\mathrm{syn}$ and $d_\mathrm{syn}^-$ indicate the depths in the synthetic spectra with the target line included and excluded.
The ratio becomes 0 if the target line explains the total absorption completely (i.e., free of blends). 
We selected the line pairs, as those \textit{unblended}, whose $d_\mathrm{syn}^- / d_\mathrm{syn}$ does not reach 0.3 at any of the above $(\Teff, \logg)$.
We present the measurements of the gravity effect on such 60 \textit{unblended} LDRs and discuss its cause in Section~\ref{sec:shift}.
In contrast, we found that the following 21 pairs include line(s) with significant blends: (T2), (T3), (T4), (T6), (T7), (T8), (T14), (T17), (T18), (T25), (T35), (T45), (T46), (T49), (T50), (T52), (T55), (T56), (T59), (T65), and (T78).
We briefly discuss how the gravity can affect the \textit{blended} LDRs in Section~\ref{sec:blended-lines}.

\subsection{The gravity effect seen in \textit{unblended} LDRs}
\label{sec:shift}

\begin{figure}
	\centering
	\includegraphics[width=1\columnwidth]{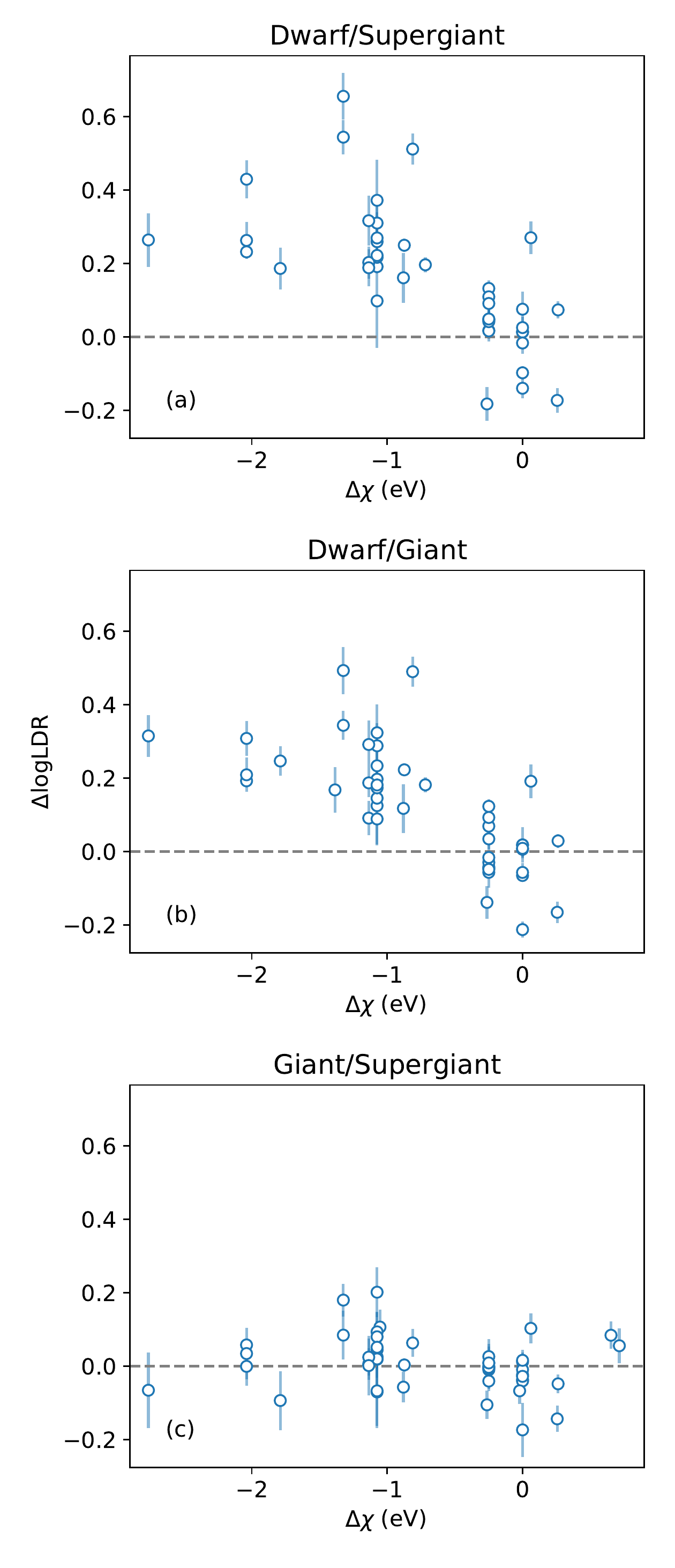}
	\caption{The $\Delta \log{\mathrm{LDR}}$ between (a) dwarf--supergiant, (b) dwarf--giant, and (c) giant--supergiant plotted against $\Delta\chi$.}
	\label{fig:separation-value}
\end{figure}

Many of the 60 line pairs without the significant blends show the offsets of LDR--$\Teff$ relations between the three groups of luminosity class.
We estimated the offsets between the LDR--$\Teff$ relations of dwarfs, giants, and supergiants for each \textit{unblended} line pair presented in T18 as follows.
Linear LDR--$\Teff$ relations were fitted separately for dwarfs, giants, and supergiants.
Then, we calculated offsets of these relations, $\Delta \log{\mathrm{LDR}}$, at $\Teff = 5000$\,K (for the dwarf--giant and dwarf--supergiant pairs) or $4500$\,K (for the giant--supergiant pairs).
The $\Delta \log{\mathrm{LDR}}$ values are listed in Table~\ref{tab:del_T} and plotted against the differences in the ionization potentials of the elements involved, $\Delta \chi \equiv \chi(\mathrm{X_{low}}) - \chi(\mathrm{X_{high}})$, in Fig.~\ref{fig:separation-value}, in which the line pairs with the $\Delta \log{\mathrm{LDR}}$ error larger than 0.2 are not included. 
The $\Delta \log{\mathrm{LDR}}$ values of the dwarf--supergiant and dwarf/giant pairs are correlated with $\Delta \chi$, but those of the giant--supergiant pairs are concentrated around zero.

\begin{figure*}
	\centering
	\includegraphics[width=2\columnwidth]{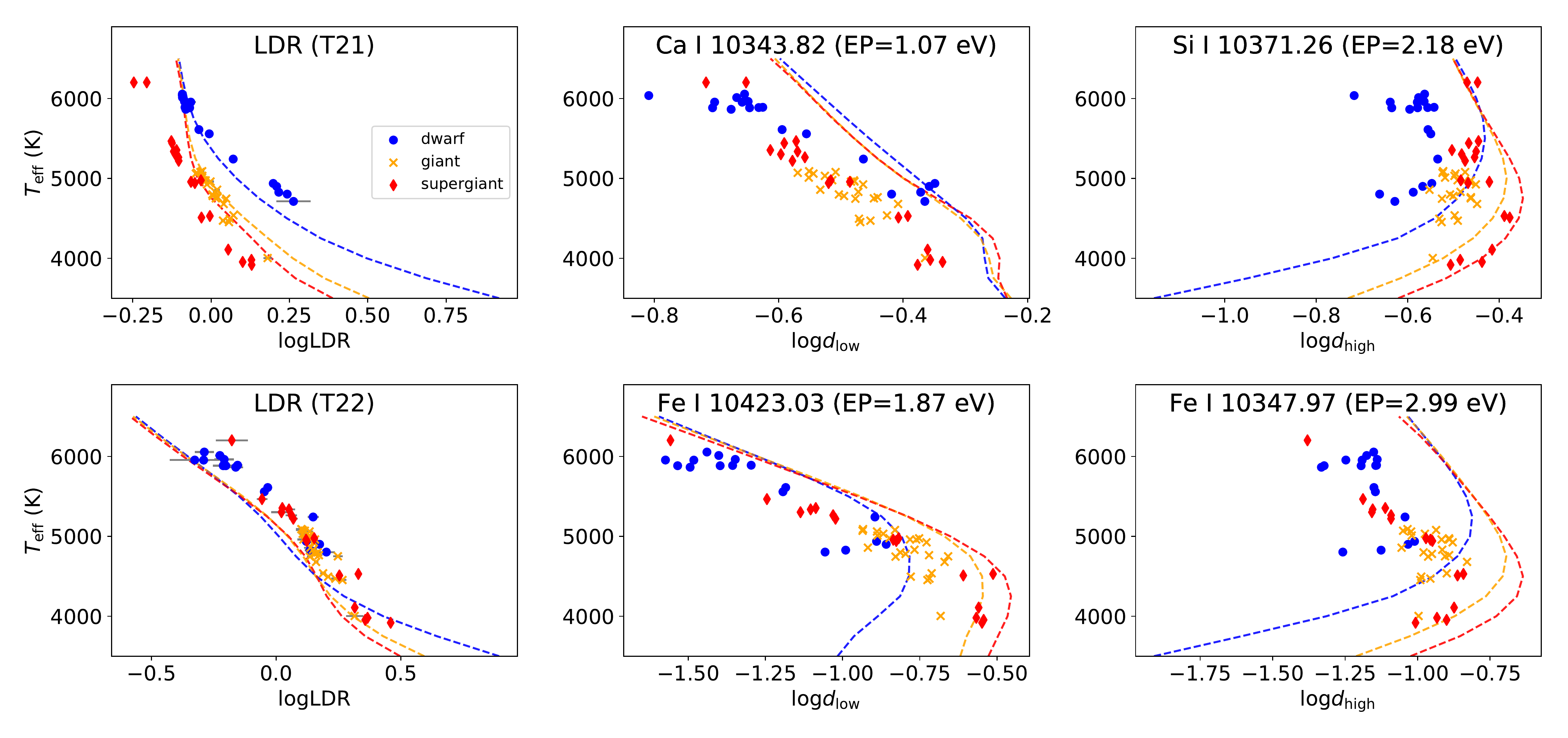}
	\caption{$\Teff$ plots against LDRs ($\log{\mathrm{LDR}}$) and line depths ($\log{d}$) for our sample stars (points) and synthetic spectra (dashed curves) for the line pairs (T21) Ca\,$\textsc{i}$ 10343.82/Si\,$\textsc{i}$ 10371.26 and (T22) Fe\,$\textsc{i}$ 10423.03/Fe\,$\textsc{i}$ 10347.97. 
	The model line depth or LDR at $\logg = 1.5$ (red curve, for supergiant), 2.5 (orange curve, for giant), and 4.5 (blue curve, for dwarf) are illustrated in each panel.}
	\label{fig:LP2122}
\end{figure*}

In order to understand how the offsets appear (or not), we use synthetic spectra to compare the trends of predicted LDRs with those measured.
Fig.~\ref{fig:LP2122} illustrates such comparisons between the measured and predicted depths/LDRs for the two line pairs, (T21) Ca\,$\textsc{i}$ 10343.82/Si\,$\textsc{i}$ 10371.26 and (T22) Fe\,$\textsc{i}$ 10423.03/Fe\,$\textsc{i}$ 10347.97, as examples.
In case of the pair (T21) which consists of Ca\,$\textsc{i}$ and Si\,$\textsc{i}$ lines, the LDR--$\Teff$ relations of giants and supergiants are close to each other, but that of dwarfs shows a clear offset.
The depth of the Si\,$\textsc{i}$ line is sensitive to $\logg$, while that of the Ca\,$\textsc{i}$ line shows almost no sensitivity.
This difference in the sensitivity to the gravity leads to the offset of the LDRs.
These trends are consistent with the theoretical curves predicted with MOOG.
We note that there are systematic offsets between the theoretical curves and the observational points, probably due to the errors in oscillator strengths, but the trends in Fig.~\ref{fig:LP2122} are consistent.
In contrast, the two lines of the pair (T22) Fe\,$\textsc{i}$ 10423.03/Fe\,$\textsc{i}$ 10347.97 are both Fe\,$\textsc{i}$ and the sensitivity of their depths to $\logg$ is predicted to be similar.
The $\Delta \log{\mathrm{LDR}}$ is negligible except at the lowest temperatures, $\Teff < 4000$\,K, and no gravity effect on the LDR was confirmed in our measurements.

\begin{table*}
    \centering
    \caption{The offsets, $\Delta \log{\mathrm{LDR}}$, between the LDR--$\Teff$ relations of the dwarf--supergiant (ds), dwarf--giant (dg), and giant--supergiant (gs) pairs.
    The 60 line pairs without significant line blend (see text) are considered here.
    The offsets were measured at $\Teff = 5000$\,K for the ds and dg pairs or at 4500\,K for the gs pair.
    For each line pair, the ID is adopted from T18, and $\Delta \chi$ indicates the difference of the ionization potentials.
    The full table is available as Supporting Information.}    
    \begin{tabular}{ccccccc}
        \hline
        ID & $\mathrm{X_{low}}$ & $\mathrm{X_{high}}$ &$\Delta\chi$\,(eV) &$\Delta \log{\mathrm{LDR_{ds}}}$ & $\Delta \log{\mathrm{LDR_{dg}}}$ & $\Delta \log{\mathrm{LDR_{gs}}}$ \\
        \hline
   (T1) &               Ti\,$\textsc{i}$ &                Fe\,$\textsc{i}$ &   -1.07 &                                               -- &                                               -- &                                  $0.05 \pm 0.10$ \\
   (T5) &               Fe\,$\textsc{i}$ &                Fe\,$\textsc{i}$ &    0.00 &                                  $0.08 \pm 0.05$ &                                  $0.02 \pm 0.05$ &                                  $0.01 \pm 0.03$ \\
   (T9) &               Ti\,$\textsc{i}$ &                Fe\,$\textsc{i}$ &   -1.07 &                                  $0.31 \pm 0.03$ &                                  $0.32 \pm 0.03$ &                                  $0.04 \pm 0.03$ \\
  (T10) &               Ti\,$\textsc{i}$ &                Si\,$\textsc{i}$ &   -1.32 &                                  $0.65 \pm 0.06$ &                                  $0.49 \pm 0.06$ &                                  $0.18 \pm 0.04$ \\
  (T11) &               Ti\,$\textsc{i}$ &                Fe\,$\textsc{i}$ &   -1.07 &                                               -- &                                               -- &                                 $-0.07 \pm 0.10$ \\
  (T12) &               Fe\,$\textsc{i}$ &                Fe\,$\textsc{i}$ &    0.00 &                                 $-0.14 \pm 0.03$ &                                 $-0.21 \pm 0.02$ &                                 $-0.02 \pm 0.03$ \\
  (T13) &               Fe\,$\textsc{i}$ &                Ni\,$\textsc{i}$ &    0.26 &                                  $0.07 \pm 0.02$ &                                  $0.03 \pm 0.02$ &                                 $-0.05 \pm 0.03$ \\
  (T15) &               Ca\,$\textsc{i}$ &                Fe\,$\textsc{i}$ &   -1.79 &                                  $0.19 \pm 0.06$ &                                  $0.25 \pm 0.04$ &                                 $-0.09 \pm 0.08$ \\
  (T16) &               Fe\,$\textsc{i}$ &                Si\,$\textsc{i}$ &   -0.25 &                                  $0.13 \pm 0.02$ &                                  $0.07 \pm 0.02$ &                                 $-0.00 \pm 0.02$ \\
  (T19) &               Fe\,$\textsc{i}$ &                Fe\,$\textsc{i}$ &    0.00 &                                 $-0.10 \pm 0.02$ &                                 $-0.07 \pm 0.02$ &                                 $-0.04 \pm 0.03$ \\
    \hline
    \end{tabular}
    \label{tab:del_T}
\end{table*}

To explain the gravity effect on the LDRs, in particular the trends seen in Fig.~\ref{fig:separation-value}, we consider how the line depths vary with $\logg$ based on a simple theoretical discussion.
\citet{Grayoasp} explains the effect of $\logg$ on the line depths as follows. 
The depth of a weak line is given as
\begin{equation}
d = \frac{F_c - F_\nu}{F_c} \propto \frac{l_\nu}{\kappa_\nu} 
\end{equation}
where $F_c$ and $F_\nu$ are the flux at the continuum level and the flux at the centre of the line, $l_\nu$ is the line absorption coefficient, and $\kappa_\nu$ is the continuum absorption coefficient.
The absorption coefficients and $\logg$ are connected through the relation between $P_\mathrm{e}$ and $\logg$, $\log{P_\mathrm{e}} \approx n\log{g}$, where the coefficient $n$ is larger than 0 but varies with temperature.
The dependencies of $l_\nu$ and $\kappa_\nu$ on the electron pressure, $\Pe$, vary with the temperature at the line formation region.
We note that all the lines adopted in T18 are neutral lines.
In the low-temperature range where most of the atoms are in the neutral state (case L1), $l_\nu$ for the neutral line is insensitive to $P_\mathrm{e}$. 
In contrast, $l_\nu \propto P_\mathrm{e}$ in the high-temperature range where most of the atoms are in the single ionization state (case L2).
For example, where Fe\,$\textsc{i}$ lines are concerned, neutral Fe atoms are dominant in the line formation region in the L1 case, but Fe$^+$ ions are dominant in the L2 case.
On the other hand, $\kappa_\nu$ is proportional to $\Pe$ where negative hydrogen (H$^-$) dominates the continuum absorption in the low-temperature range (case C1), whereas $\kappa_\nu$ is insensitive to $\Pe$ where neutral hydrogen (H) dominates in the high-temperature range (case C2).

Combining the different cases of the line and continuum absorption coefficients, the following three situations may appear for our target lines at $\Teff$ around the range of 3500--8000\,K:

\begin{enumerate}
    \item L1--C1: $\log{l_\nu} \approx \mathrm{constant}$ and $ \log{\kappa_\nu} \approx n \logg$, leading to $\log{d} \approx -n\logg$
    \item L2--C1: $\log{l_\nu} \approx n\logg$ and $\log{\kappa_\nu} \approx n\logg$, leading to $\log{d} \approx \mathrm{constant}$
    \item L2--C2: $\log{l_\nu} \approx n\logg$ and $\log{\kappa_\nu} \approx \mathrm{constant}$, leading to $\log{d} \approx n\logg$
\end{enumerate}

For some of the elements we consider, most atoms are neutral in cool stars, but get ionized at high effective temperatures. 
No metals relevant here remain neutral-dominated at $\Teff > 7000$\,K.
The main source of the continuum absorption changes from $\mathrm{H^-}$ to $\mathrm{H}$ at $\Teff$ around 7000\,K; therefore we do not expect the presence of the L1--C2 case.
As a result, the L--C cases, i.e., the sensitivity of the line depth to $\logg$ as listed above, changes from L1--C1 to L2--C1 and further to the L2--C2 case with increasing $\Teff$ if most of the atoms are neutral in the lowest $\Teff$.
If the atoms in the atmosphere are ionized even at $\Teff \approx 3500$\,K, the L1--C1 case does not appear. 
We note that the maximum $\Teff$ in our sample is 6202\,K, and thus no line is expected to be in the L2--C2 case within the range of our observational data.
If the low-EP line is in the L2--C1 case and the high-EP line is in the L1--C1 case, $\logLDR = \log{d_\mathrm{low}} - \log{d_\mathrm{high}} \approx n\logg$, but $\logLDR$ does not depend on $\logg$ if both the lines are in the L2--C1 case.

Let us consider the ionization degree $R_\mathrm{ion} = N_1 / (N_0 + N_1)$, where $N_0$ and $N_1$ indicate the densities of the neutral and singly ionized atoms and the higher ionization states are not taken into account.
This degree is given by the Saha ionization equation
\begin{equation}
    \frac{N_1}{N_0}=\frac{(2\pi m_\mathrm{e})^{3/2} (kT)^{5/2}}{h^3} \frac{2u_1(T)}{u_0(T)} \frac{e^{-\chi/kT}}{P_e},   
\end{equation}
with the temperature $T$, the ratios of the partition functions $\frac{u_1(T)}{u_0(T)}$, the electron pressure $P_e$, and the ionization potential $\chi$, while $\pi$, $m_e$, $k$, and $h$ are the constants in the usual notation (see \citealt{Grayoasp}). 
We calculated the $R_\mathrm{ion}$ index of each line in case of the stellar atmosphere for $\Teff=5000$\,K and $\logg=4.0$\,dex based on the temperature and electron pressure in the line-forming region given by MOOG.
The $R_\mathrm{ion}$ of the Na, Ca, Cr, Ti, and Mg lines at $\Teff = 5000$\,K and $\logg = 4.0$ are around 0.9 (mostly ionized), while those of Al, Mn, Ni, Co, Fe, and Si are smaller than 0.8 (not fully ionized) and decrease as $\chi$ increases except for Al.
These two groups of elements, the group (a) such as Ca having lower $\chi$ and the group (b) such as Fe having higher $\chi$, can be separated by an ionization potential of 7.5\,eV except for Mg and Al.
The $\chi$ of Mg is 7.6\,eV, but its $R_\mathrm{ion}$ is close to 1 because its ratio of the partition functions is very large ($\approx 300$) at $\Teff = 5000$\,K.
The $\chi$ of Al, in contrast, is 6.1\,eV, whereas its $R_\mathrm{ion}$ is around 0.3 since its ratio of partition function is small (around 0.001).
For other elements, their ratios of the partition functions are close to 1.

Then, the L--C cases of the lines in the group (a) change from L1--C1 to L2--C1 as $\Teff$ increases, and the lines in the group (b) remain in the L2--C1 case in the $\Teff$ range we consider. 
When a line pair consists of one line from the group (a) and another one from the group (b), it tends to have a large $\Delta \chi$ and a large $\logg$ effect (i.e., a large $\Delta \log{\mathrm{LDR}}$ value).
On the contrary, if both lines of a line pair are included in the same group, both $\Delta \chi$ and the $\logg$ effect tend to be small.
Furthermore, $R_\mathrm{ion}$ becomes larger as $\logg$ decrease according to the Saha ionization equation since $\log{P_\mathrm{e}} \approx n\logg$.
The $R_\mathrm{ion}$ of all the elements in giants and supergiants are closer to 1 compared to those in case of dwarfs; thus the $\logg$ effect between giants and supergiants is smaller than that between dwarfs and giants or supergiants.

The gravity effects on the line depths and on the LDR are consistent with the theoretical expectation described above, e.g., for the line pairs (T21) Ca\,$\textsc{i}$ 10343.82/Si\,$\textsc{i}$ 10371.26 and (T22) Fe\,$\textsc{i}$ 10423.03/Fe\,$\textsc{i}$ 10347.97 as illustrated in Fig.~\ref{fig:LP2122}.
The Ca\,$\textsc{i}$ line is insensitive to $\logg$ (case L2--C1) across the $\Teff$ range in consideration.
For the Si\,$\textsc{i}$ and Fe\,$\textsc{i}$ lines, the $\log{d}$ sensitivity to $\logg$ is negative between 3500 and 5500~K (case L1--C1), but almost negligible at $\Teff$ higher than 5500~K (case L2--C1).
Thus the difference in the line-depth sensitivity to $\logg$ following the ionization of the elements leads to the gravity effect on the line pair (T21) Ca\,$\textsc{i}$ 10343.82/Si\,$\textsc{i}$ 10371.26, and other line pairs that show the systematic shifts of the LDR relations.
For the line pairs with no significant shift, the situation is similar to the line pair (T22) Fe\,$\textsc{i}$ 10423.03/Fe\,$\textsc{i}$ 10347.97, where the sensitivity of the two lines in each line pair to $\logg$ cancel out.

The gravity effect is not strong enough to be significant inside each luminosity class.
For example, \citet{biazzo_effective_2007} reported the gravity effect considering the sample of dwarfs and giants with a range of luminosity (or surface gravity).
Their figure 5 suggests that the effect on LDR is as small as $\pm 0.025$ for dwarfs in the range of $-0.35 < \log{L/L_\mathrm{ZAMS}} < 0.8$, where $L_\mathrm{ZAMS}$ indicates the luminosity of zero-age main sequence, and it corresponds to $\pm 45$\,K or smaller considering the slopes of LDR--$\Teff$ relations, at $\Teff = 5000$\,K, according to table 3 in \citet{biazzo_effective_2007}.
The expected effect is thus significantly smaller than the scatter of the LDR--$\Teff$ relations we obtained.
In addition, we calculated the $\Delta \log{\mathrm{LDR}}$ values of the T18 line pairs in $\Teff=4500$\,K for dwarfs (between $\logg=4.0$ and $4.5$\,dex) and supergiants (between $\logg=1.0$ and $2.0$\,dex) using synthetic spectra.
The mean $\Delta \log{\mathrm{LDR}}$ value is $\pm 0.02$, similar in both groups, leading to the gravity effect of $\pm 40$\,K or smaller in $\Teff$.
It is again smaller than the scatter of the LDR--$\Teff$ relations.
To minimize the gravity effect on $\Teff$ based on the LDR method and keep the form of LDR--$\Teff$ relation simple, we consider the LDR--$\Teff$ relations calibrated separately for dwarfs and supergiants in Section~\ref{sec:new-relations}, but it is unnecessary to add gravity-dependent terms to the relations for each luminosity class.

\subsection{The gravity effect seen in \textit{blended} LDRs}
\label{sec:blended-lines}

There are 21 line pairs with at least one line blended by other line(s) significantly at $\Teff = 4500$ and/or $5000$\,K.
If the target line is blended by a line with no or little gravity effect, the gravity effect on the composite absorption tends to be similar or smaller compared to the effect on the target line itself.
On the contrary, if the blending line is sensitive to the gravity, the total gravity effect is different from that of the target line.
We created the synthetic spectra using the target line blended with atomic or molecular lines with the stellar parameters $\Teff=4000$ and $5000$\,K, and $\logg=1.0$, $2.0$, $3.0$, and $4.0$\,dex.
The target lines are mainly blended by atomic lines at $\Teff = 5000$\,K, while the molecular CN and CH lines become strong at $\Teff = 4000$\,K.
The blending lines of the line pairs (T2), (T3), (T8), (T25), (T35), (T45), (T49), (T50), and (T78) are not sensitive to the gravity, and thus the gravity effects on these line pairs are mainly from the target lines themselves. 
The blending lines for the other line pairs are sensitive to the gravity, and thus the sensitivity of the blending lines tends to be more important.
If the blending line is atomic (whether neutral or ionic), the gravity effect can be understood by considering the three L--C cases, (i) to (iii), discussed in Section~\ref{sec:shift} for the blending line together with the LDR line pair.
Ionic lines, incidentally, is in the L1--C1 case throughout the $\Teff$ range we consider.
To characterize the behaviour of molecular lines, we explored the trends of CH, CN, and CO lines in synthetic spectra and found that they get stronger with decreasing $\logg$ at $\Teff < 4750$\,K where the contamination of molecular lines becomes significant. 
The trends for CN and CO lines are  consistent with some previous results,  e.g., \citet{Kleinmann1986} and \citet{ Lancon2007}. 
This behaviour is similar to that of an atomic line in the L1--C1 case discussed above and thus lead to the gravity effect accordingly.

It is worthwhile to note that the blending has another side effect on the LDR--$\Teff$ relations: the sensitivity to spectral resolution.
We tested how changing the resolution shifts, or not, the relations using synthetic spectra.
The relations of line pairs with no blending are independent of the resolution as expected, since the changes in the depths of low- and high-EP lines cancel each other.
However, when the blending into the LDR lines are significant (especially if larger than 30\% in depth), the relations are subject to offsets caused by changing the spectral resolution.
When blended, the line depth tends to be relatively larger with decreasing the resolution because the contribution of the blending line(s) gets larger.
Thus, for example, the LDR with the low-EP line more blended becomes larger with decreasing resolution.
Characterizing the effects of the blending which depend on the separation between the target line and contaminating lines and the ratio of their strengths is not simple. 
It is therefore desirable to calibrate the LDR--$\Teff$ relations, again, for different spectrographs providing different spectral resolutions.

\section{LDR--$\Teff$ relations in $YJ$ band for dwarfs and supergiants}
\label{sec:new-relations}

The surface gravity not only gives offsets to some LDRs.
Useful sets of lines with significant depths also change with the surface gravity even for the same range of $\Teff$.
Therefore, we decided to search for new line pairs that are useful for dwarfs and supergiants and calibrate their LDR--$\Teff$ relations.
Some lines used in T18 may be included in our new line pairs, but they do not necessary appear in the same line pairs as those in T18.

A set of relatively isolated spectral lines were identified and selected as follows.
We first selected the absorption features whose depths are smaller than 0.5 in all the observed spectra in each group of dwarfs and supergiants.
The lines with depth larger than 0.5 may be strongly influenced by various parameters such as damping constants \citep{Kovtyukh2006a}, and thus they were excluded.
Then, to identify individual absorption lines, we used the model spectra synthesized with the typical parameters for each group: $\Teff = 5500$\,K, $\logg = 4.5$\,dex, and $\FeH = 0.0$\,dex for dwarfs, and $\Teff = 5500$\,K, $\logg = 2.0$\,dex, and $\FeH = 0.0$\,dex for supergiants.
All the lines in the VALD3 line list around the wavelength of each selected feature were included in the synthetic spectra, and the atomic line that has the largest contribution to the feature was identified.
We did not consider ion lines and atomic lines of C, N, or O as candidates for LDRs, because those lines are expected to depend very strongly on $\logg$ (see e.g., section~2 of \citealt{Kovtyukh2003} and section~3 of \citealt{Kovtyukh2007}).
We also rejected lines with the ratio $d_\mathrm{syn}^- / d_\mathrm{syn}$ (defined in Section~\ref{sec:gravity-effect}) larger than 0.3.
Thus, 150 (for dwarfs) and 223 (for supergiants) lines were selected to be used for the line pair selection in the next step. 

We followed a procedure of the line pair selection similar to the one described in section 3.3 of T18 to find a set of line pairs that gives precise temperatures through tight LDR relations.
We considered the combinations of the line pairs that meet the following criteria:
(1)~Each line is used for up to one line pair only. 
(2)~The two lines of each pair are included within the same order, and the difference in EP of the two lines should be larger than 1\,eV.
(3)~The line pair candidates are fitted by the linear relation $\Teff = a \log{\mathrm{LDR}} + b$, and every selected relation must have a negative slope ($a < 0$), the residual smaller than 150\,K, and 10 or more stars useful for the fitting.
During this step, all the LDRs for which we accepted the depths of both lines (see Section~\ref{sec:LDR-measure}) were used for the fitting.
In order to take together into account the errors in both the LDRs and those in the reference $\Teff$, we performed the orthogonal distance regression \citep{boggs_rogers_1989}.
Then, for each set of line pairs, $M_k$, the evaluation function defined as $E(M_k;e) = E_T(M_k) + eE_\lambda(M_k)$ was calculated and used to evaluate the goodness of the line pair set.
The evaluation function consists of two parts: $E_T(M_k)$ is the standard deviation of the difference between the weighted mean $\Teff$ determined by the linear functions and the reference $\Teff$, while $E_\lambda(M_k)$ is the root mean square of the wavelength separation of the paired lines (see also, the definitions in T18).
The coefficient $e$ controls the tolerance of the wavelength separation, and different sets of line pairs may be chosen with different $e$ values.
We adopted $e = 0.5$ from T18 and selected the line pair set that gives the smallest $E(M_k; e)$ value.

\begin{figure*}
	\centering
	\includegraphics[width=2\columnwidth]{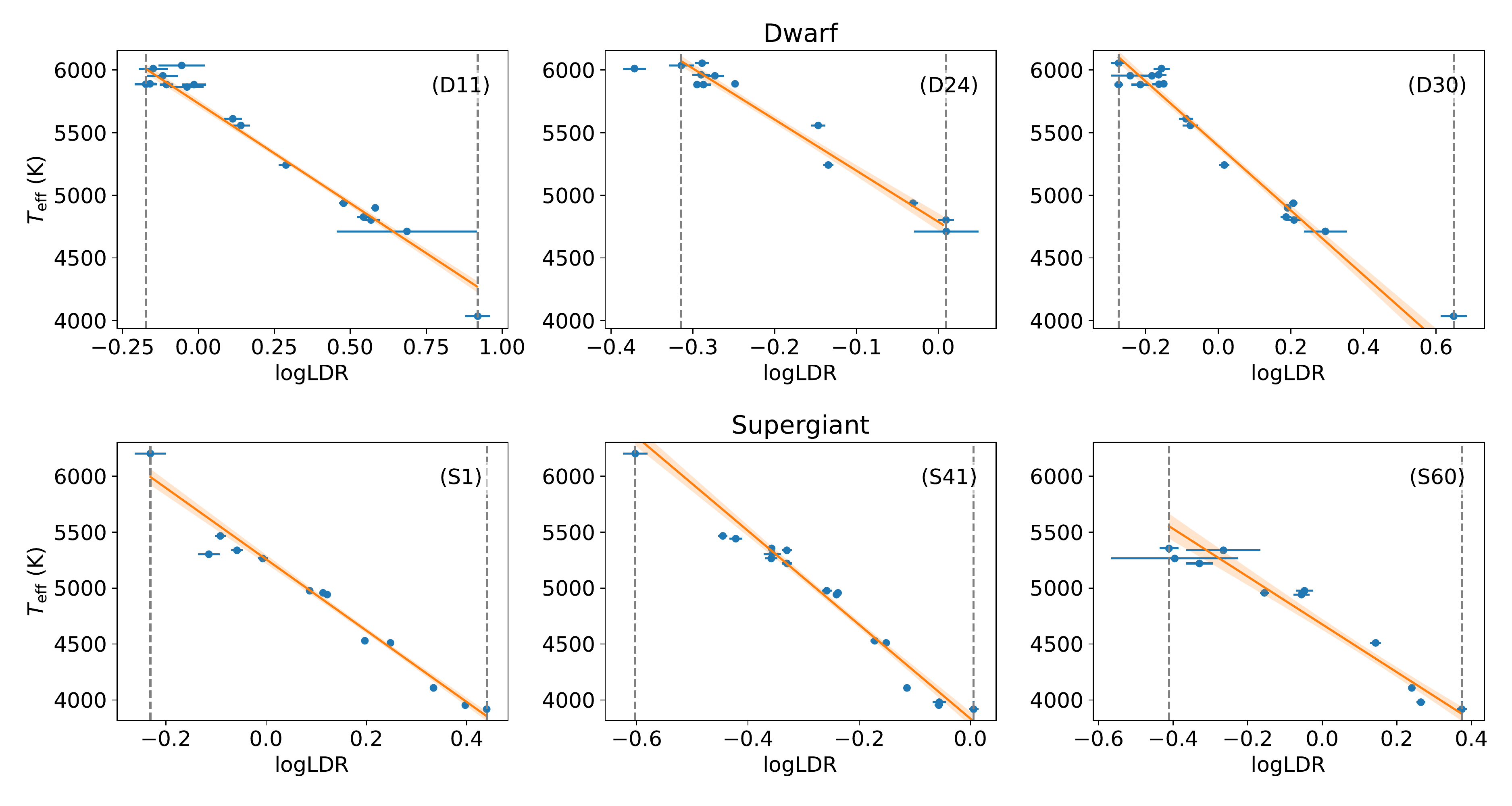}
	\caption{Examples of the LDR--$\Teff$ relations for dwarfs and supergiants. The IDs of the line pairs (given in Table~\ref{tab:linelist_d} and \ref{tab:linelist_spg}) are labelled in the upper-right corner. The vertical dashed lines indicate the $\log{\mathrm{LDR}}$ range of each relation. Similar plots for all the line pairs are available as Supporting Information.}
	\label{fig:LP_example_dwarf}
\end{figure*}

Some of the relations after the line pair selection show deviation from the linear correlation towards the high- or low-temperature end, and we excluded the non-linear trends by limiting the range of $\log{\mathrm{LDR}}$.
Although \citet{Jian2019a} used higher order functions, only $\sim$\,15 stars are available for calibrating each LDR--$\Teff$ relation in this work, and limiting the $\log{\mathrm{LDR}}$ ranges allows more robust estimates of $\Teff$ of individual objects.
We also note that the number of our relations is large enough to give at least 18 useful LDRs for each star even if some LDRs are outside the range to be accepted, while \citet{Jian2019a} considered only 11 LDR relations from $H$-band spectra in total. 
Finally, 38 line pairs for dwarfs and 69 line pairs for supergiants were selected and fitted by the linear relation. 
Among the 162 lines (of 81 pairs) used in T18, 45 and 84 lines for dwarfs and supergiants, respectively, are included in the sets of line pairs we selected, but only 7 and 8 line pairs are common with those selected for giants in T18.
Table~\ref{tab:linelist_summary} summarizes the wavelength range of each order and the numbers of the line pairs as well as those in T18 for comparison, while Table~\ref{tab:linelist_d} and  \ref{tab:linelist_spg} list the line pairs and their LDR--$\Teff$ relations for dwarfs and supergiants respectively.
The ID of these line pairs for dwarfs and supergiants are prefixed by `D' or `S', respectively.
The coefficients and residual around the fit are listed together with the number of the data points and the $\log{\mathrm{LDR}}$ range used for obtaining each relation in these tables.
The parameters of the LDR--$\Teff$ relations in Table~\ref{tab:linelist_d} and \ref{tab:linelist_spg} are obtained with only LDRs within the given $\log{\mathrm{LDR}}$ ranges, whereas Fig.~\ref{fig:LP_example_dwarf} plots the data points including those outside the $\log{\mathrm{LDR}}$ ranges.

\begin{table}
	\centering
	\caption{Numbers of the selected line pairs in individual order, whose wavelength ranges $\lambda_\mathrm{min} < \lambda < \lambda_\mathrm{max}$, for dwarfs, giants (adopted from T18), and supergiants. 
	The wavelength range of each order is from WINERED website (\url{http://merlot.kyoto-su.ac.jp/LIH/WINERED}).}
	\begin{tabular}{ c c c c c c }
		\hline
    Order &  $\lambda_\mathrm{min}$ &  $\lambda_\mathrm{max}$ & \multicolumn{3}{c}{$N_\mathrm{pair}$}  \\ 
     & ($\mu$m) & ($\mu$m) & Dwarf & Giant & Supergiant \\ 
    \hline
       57 &                   0.976 &                   0.992 &                         3 &                               1 &                              4 \\
       56 &                   0.992 &                   1.010 &                         1 &                              10 &                              6 \\
       55 &                   1.010 &                   1.028 &                         2 &                               4 &                              4 \\
       54 &                   1.028 &                   1.048 &                         5 &                              10 &                             10 \\
       53 &                   1.048 &                   1.068 &                         7 &                               8 &                              8 \\
       52 &                   1.068 &                   1.089 &                         7 &                              14 &                             13 \\
       48 &                   1.156 &                   1.180 &                         2 &                               3 &                              1 \\
       47 &                   1.180 &                   1.205 &                         2 &                               3 &                              1 \\
       46 &                   1.205 &                   1.232 &                       0  &                               4 &                              3 \\
       45 &                   1.232 &                   1.260 &                         3 &                               5 &                              7 \\
       44 &                   1.260 &                   1.290 &                       0 &                               7 &                              6 \\
       43 &                   1.290 &                   1.319 &                         5 &                              12 &                              8 \\
       \hline
 Total &                   0.976 &                   1.319 &                        37 &                              81 &                             71 \\

		\hline
	\end{tabular}
	\label{tab:linelist_summary}
\end{table}

\begin{table*}
	\centering
	\caption{First 10 line pairs in the line pair set of our dwarf LDR--$\Teff$ relations. 
	Listed for each line pair are the information of low- and high-EP lines (species, air wavelength $\lambda$, and EP taken from VALD3) as well as the coefficients (slope $a$ and intercept $b$), the residual of the fitted relation ($\sigma$), the number of stars ($N$), and the $\log{\mathrm{LDR}}$ range used for calibrating the relation. 
	The full list of the 38 line pairs is available as Supporting Information.}
	\begin{tabular}{ c c c c c c c c c c c c c c}
		\hline
		& &\multicolumn{3}{c}{Low-excitation line} & \multicolumn{3}{c}{High-excitation line} & \multicolumn{5}{c}{LDR--$\Teff$ relation}\\
   ID &  Order & $\mathrm{X_{low}}$ &   $\lambda$ (\AA) &  EP (eV) & $\mathrm{X_{high}}$ &   $\lambda$ (\AA) &  EP (eV) &    $a$ &   $b$ & $\sigma$ (K) &   $N$ &             $\log{\mathrm{LDR}}$ range \\\hline
  (D1) &    57 &               Ti\,$\textsc{i}$  &         9770.301 &  0.8484 &                Fe\,$\textsc{i}$  &         9861.734 &  5.0638 &  -2347 &  5201 &          116 &  13 &   $[-0.39$:$0.49]$ \\
  (D2) &    57 &               Ti\,$\textsc{i}$  &         9787.687 &  0.8259 &                Si\,$\textsc{i}$  &         9887.047 &  6.2227 &  -1808 &  6189 &          134 &  12 &    $[0.08$:$0.86]$ \\
  (D3) &    56 &               Ti\,$\textsc{i}$  &        10034.491 &  1.4601 &                Si\,$\textsc{i}$  &        10068.329 &  6.0986 &   -859 &  5268 &           59 &  12 &   $[-0.85$:$0.59]$ \\
  (D4) &    54 &               Fe\,$\textsc{i}$  &        10340.885 &  2.1979 &                Fe\,$\textsc{i}$  &        10469.652 &  3.8835 &  -6356 &  4256 &          111 &  18 &   $[-0.29$:$0.02]$ \\
  (D5) &    54 &               Ca\,$\textsc{i}$  &        10343.819 &  2.9325 &                Si\,$\textsc{i}$  &        10371.263 &  4.9296 &  -3634 &  5615 &           73 &  20 &   $[-0.09$:$0.52]$ \\
  (D6) &    54 &               Fe\,$\textsc{i}$  &        10395.794 &  2.1759 &                Si\,$\textsc{i}$  &        10407.037 &  6.6161 &  -1811 &  6258 &           61 &  16 &    $[0.09$:$1.03]$ \\
  (D7) &    54 &               Ti\,$\textsc{i}$  &        10396.802 &  0.8484 &                Ni\,$\textsc{i}$  &        10330.228 &  4.1054 &  -1660 &  5733 &           76 &  20 &   $[-0.24$:$0.90]$ \\
  (D8) &    54 &               Fe\,$\textsc{i}$  &        10423.027 &  2.6924 &                Fe\,$\textsc{i}$  &        10347.965 &  5.3933 &  -2802 &  5360 &          121 &  18 &   $[-0.33$:$0.20]$ \\
  (D9) &    54 &               Fe\,$\textsc{i}$  &        10423.743 &  3.0713 &                Si\,$\textsc{i}$  &        10288.944 &  4.9201 &  -2483 &  4817 &          127 &  19 &   $[-0.54$:$0.29]$ \\
 (D10) &    53 &               Cr\,$\textsc{i}$  &        10486.250 &  3.0106 &                Si\,$\textsc{i}$  &        10603.425 &  4.9296 &  -2364 &  4043 &          130 &  15 &  $[-0.87$:$-0.08]$ \\
		\hline
	\end{tabular}
	\label{tab:linelist_d}
\end{table*}

\begin{table*}
	\centering
	\caption{Same as Table~\ref{tab:linelist_d} but for supergiants. The full list of the 69 line pairs is available as Supporting Information.}
	\begin{tabular}{ c c c c c c c c c c c c c c}
		\hline
		& &\multicolumn{3}{c}{Low-excitation line} & \multicolumn{3}{c}{High-excitation line} & \multicolumn{5}{c}{LDR--$\Teff$ relation}\\
   ID &  Order & $\mathrm{X_{low}}$ &   $\lambda$ (\AA) &  EP (eV) & $\mathrm{X_{high}}$ &   $\lambda$ (\AA) &  EP (eV) &    $a$ &   $b$ & $\sigma$ (K) &   $N$ &             $\log{\mathrm{LDR}}$ range \\\hline
  (S1) &    57 &               Ti\,$\textsc{i}$  &         9783.311 &  0.8360 &                Fe\,$\textsc{i}$  &         9861.734 &  5.0638 &  -3189 &  5257 &           83 &  13 &   $[-0.23$:$0.44]$ \\
  (S2) &    57 &               Ti\,$\textsc{i}$  &         9787.687 &  0.8259 &                Fe\,$\textsc{i}$  &         9868.186 &  5.0856 &  -2993 &  5638 &          104 &  17 &   $[-0.13$:$0.65]$ \\
  (S3) &    57 &               Ti\,$\textsc{i}$  &         9879.583 &  1.8732 &                Fe\,$\textsc{i}$  &         9811.504 &  5.0117 &   -862 &  4003 &          114 &  12 &   $[-1.55$:$0.33]$ \\
  (S4) &    56 &               Ti\,$\textsc{i}$  &         9927.351 &  1.8792 &                Fe\,$\textsc{i}$  &         9980.463 &  5.0331 &  -2022 &  4887 &          126 &  13 &   $[-0.37$:$0.49]$ \\
  (S5) &    56 &               Fe\,$\textsc{i}$  &         9944.207 &  5.0117 &                Si\,$\textsc{i}$  &        10068.329 &  6.0986 &  -3033 &  5236 &          120 &  17 &   $[-0.44$:$0.44]$ \\
  (S6) &    56 &               Fe\,$\textsc{i}$  &         9987.868 &  2.1759 &                Mg\,$\textsc{i}$  &         9986.475 &  5.9320 &  -2614 &  3967 &          100 &  14 &   $[-0.93$:$0.01]$ \\
  (S7) &    56 &               Ti\,$\textsc{i}$  &        10011.744 &  2.1535 &                Na\,$\textsc{i}$  &         9961.256 &  3.6170 &  -3063 &  4403 &           87 &  11 &   $[-0.36$:$0.23]$ \\
  (S8) &    56 &               Ti\,$\textsc{i}$  &        10059.904 &  1.4298 &                Fe\,$\textsc{i}$  &        10041.472 &  5.0117 &  -1161 &  4695 &           77 &  14 &   $[-0.79$:$0.60]$ \\
  (S9) &    56 &               Fe\,$\textsc{i}$  &        10081.393 &  2.4242 &                Mg\,$\textsc{i}$  &         9993.209 &  5.9328 &  -2182 &  4207 &           87 &  14 &   $[-0.58$:$0.09]$ \\
 (S10) &    55 &               Fe\,$\textsc{i}$  &        10155.162 &  2.1759 &                Ni\,$\textsc{i}$  &        10193.224 &  4.0893 &  -2705 &  4535 &          123 &  16 &   $[-0.47$:$0.29]$ \\
\hline
	\end{tabular}
	\label{tab:linelist_spg}
\end{table*}

\begin{figure*}
	\centering
	\includegraphics[width=2\columnwidth]{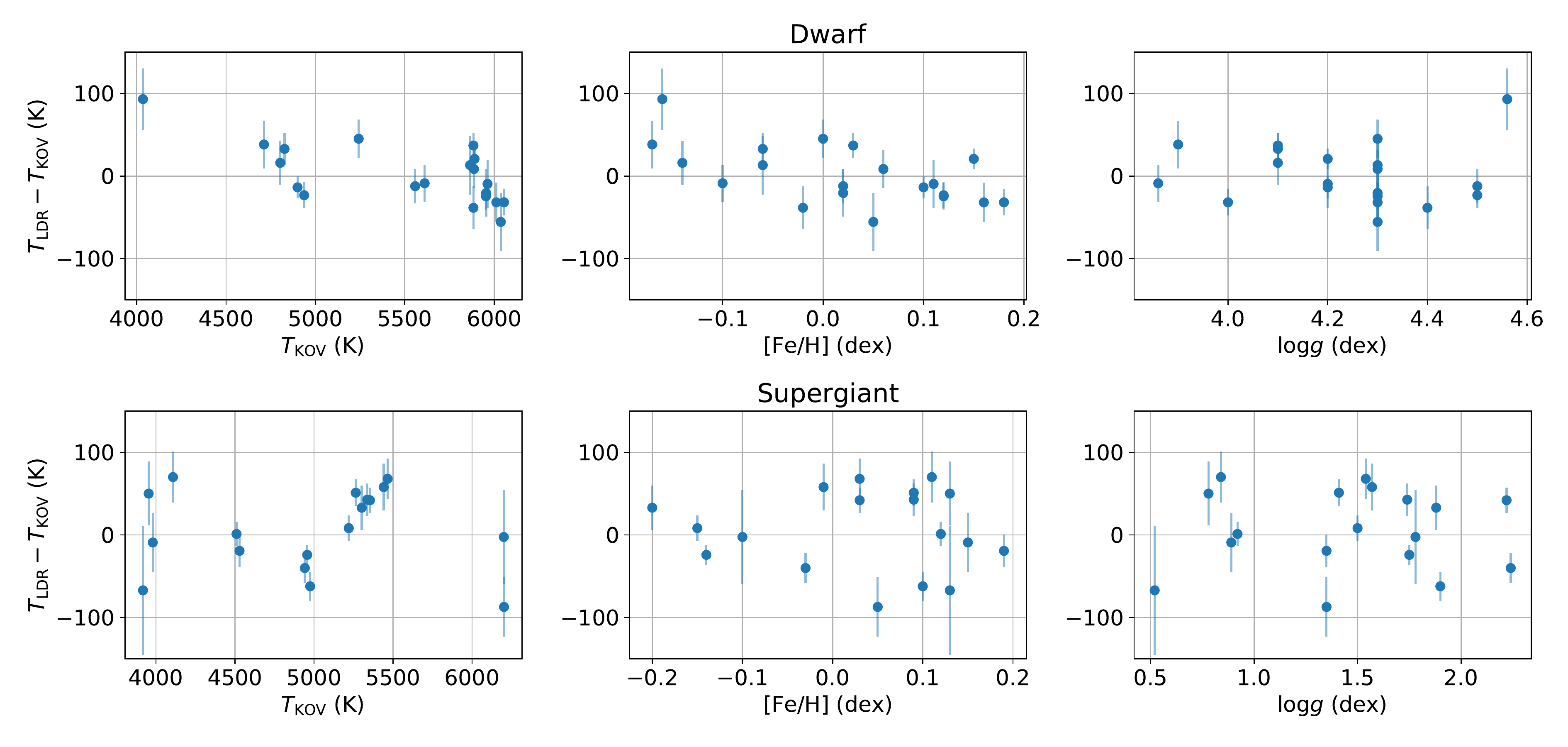}
	\caption{Plots of $\Tldr - \Tkov$ against $\Tkov$, $\FeH$, and $\logg$ for dwarfs and supergiants.}
	\label{fig:comp_dwarf}
\end{figure*}

\begin{figure}
	\centering
	\includegraphics[width=\columnwidth]{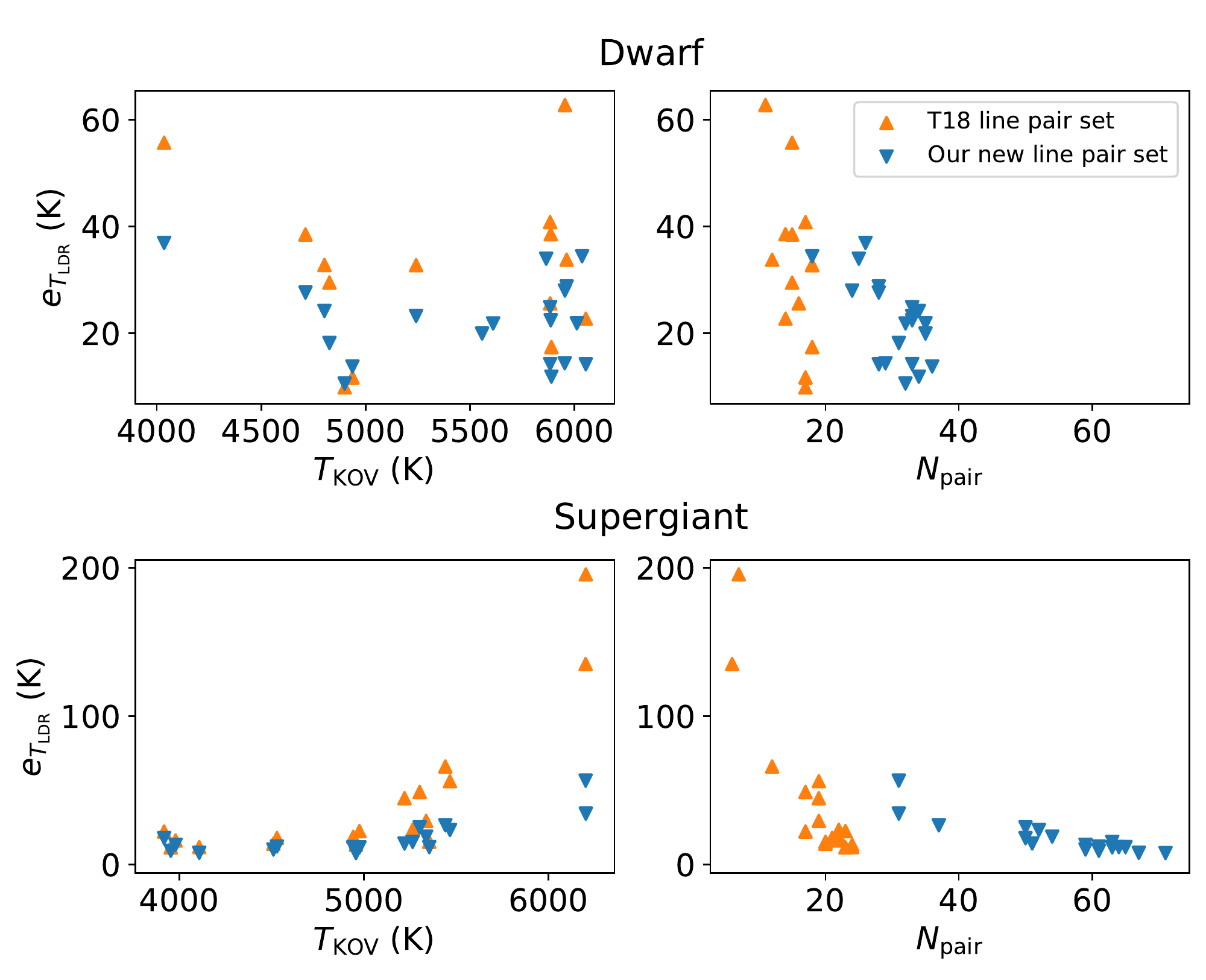}
	\caption{Comparison of $e_{T_\mathrm{LDR}}$ calculated using the T18 line pairs and our new line pairs plotted against $\Teff$ and $N_\mathrm{pair}$ for dwarfs and supergiants.}
	\label{fig:e_T_comp}
\end{figure}

The residual around each LDR--$\Teff$ relation ranges from 60 to 150\,K, which is similar to those in T18 for both dwarfs and supergiants.
The final temperature derived with our LDR relations for each star, $\Tldr$, is the weighted mean of the temperatures, $\Tldr_i$, calculated with all the individual LDR--$\Teff$ relations for which the LDRs were measured and within the $\log{\mathrm{LDR}}$ ranges given in Table~\ref{tab:linelist_d} or \ref{tab:linelist_spg}. 
The weight, $w_i$, is calculated by $w_i = 1 / e^2_{\Tldr_i}$ where $e_{\Tldr_i}$ is the error of $\Tldr_i$.
The error of $\Tldr$, $e_{T_\mathrm{LDR}}$, is calculated through the standard error of weighted mean,
\begin{equation}
e_{T_\mathrm{LDR}} = \sqrt{\frac{1}{N_\mathrm{pair}} \frac{\sum w_i (\Tldr_i-\Tldr)^2}{V_1-V_2/V_1}},
\end{equation}
where $V_1 = \sum w_i$, $V_2 = \sum w_i^2$, and $N_\mathrm{pair}$ is the number of available line pairs.
To compare our temperature scale with the ones in \citet{Kovtyukh2003, Kovtyukh2006a, Kovtyukh2006b} and \citet{Kovtyukh2007}, we compare $\Tldr$ with the catalogue temperatures, $\Tkov$, in Fig.~\ref{fig:comp_dwarf}.
The differences, $\Tldr - \Tkov$, are entirely within $\pm\,100$\,K, and there is almost no trend between the differences and stellar parameters.
The $e_{T_\mathrm{LDR}}$ are 10--30\,K for both dwarfs and supergiants as shown in Fig.~\ref{fig:e_T_comp}.
Some dwarfs tend to have larger $e_{T_\mathrm{LDR}}$ because the number of line pairs is smaller in many cases. 
The errors for supergiants are similar to those in T18 for giants (around 14\,K on average) in the $\Teff$ range between 4000 and 5500\,K, while those in the higher $\Teff$ range are slightly larger since $N_\mathrm{pair}$ is small. 

We also compare the performance of our new line pairs with that of the T18 line pairs for dwarfs and supergiants.
For dwarfs, only 18 line pairs from T18 satisfy the criteria of our line pair selection, and they only cover the $\Teff$ range from 4700 to 6000\,K. 
Although the $\Tldr - \Tkov$ values found with the T18 set are similar to those plotted in Fig.~\ref{fig:comp_dwarf}, the average $e_{T_\mathrm{LDR}}$, 32\,K, is slightly larger than that from our new set of line pairs for dwarfs.
For supergiants, only 24 line pairs from T18 are useful, and the resultant $\Tldr - \Tkov$ values show larger scatter and deviate from 0 towards the high-temperature end.
The average $e_{T_\mathrm{LDR}}$, 42\,K, is also larger than the average $e_{T_\mathrm{LDR}}$ found with our new set for supergiants.
Fig.~\ref{fig:e_T_comp} compares the $e_{T_\mathrm{LDR}}$ values for dwarfs and supergiants derived with the T18 line pairs and those derived with our new line pairs.
For both dwarfs and supergiants, the $e_{T_\mathrm{LDR}}$ with our line pairs are clearly smaller than those with the T18 line pairs.

Finally, we estimate the effect of abundance ratio ($[\XL/\XH]$) on the calibration of LDR--$\Teff$ relations and $\Tldr$.
Among the targets, there are 19 dwarfs and 12 supergiants having the abundance ratios measured by \citet{Luck2014, Luck2017}.
No clear trend is seen between metallicity and the abundance ratios relevant to our line pairs and we merely find that the scatter in $[\XL/\XH]$ is around $\pm 0.05$\,dex for dwarfs and $\pm 0.1$\,dex for supergiants in most cases, within the metallicity range we discuss, $-0.2 < \FeH < 0.2$\,dex.
The larger scatter for supergiants may well be attributed to the measurement errors rather than the true scatter.
The scatter for [Na/Fe] and [Mg/Fe] for supergiants are larger, $\pm 0.15$\,dex, but their impacts on our analysis are expected to be relatively small because only 2 Na and 7 Mg lines are included in our line pairs for supergiants.
The change of 0.05\,dex in $[\XL/\XH]$ would correspond to the change of 0.05 in $\logLDR$ if the two lines are within the linear region.
However, many lines used in this work are saturated at around the solar metallicity, and the changes in $\logLDR$ are thereby smaller, around $\pm 0.025$, corresponding to roughly $\pm 50$\,K in the changes, or systematic errors, of $\Teff$. 
Thus, we expect that the effect of abundance ratio contributes to the residual of each relation in an amount of around 50\,K if the scatter in $[\XL/\XH]$ of the calibrating sample is 0.05\,dex.
This effect may also bias the $\Tldr$ by at most 50\,K, if all the $[\XL/\XH]$ values of a given star are higher or lower than the solar by 0.05\,dex.
Considering this effect, combining low- and high-EP lines of the same element certainly has an advantage, and such line pairs are also insensitive to the gravity effect according to our conclusion in Section~\ref{sec:gravity-effect}. 
However, there are not so many such line pairs that show tight LDR--$\Teff$ relations.
Including line pairs of different elements, as done by us and many studies, increases the useful line pairs leading to the higher precision for stars with a limited range of abundances. 
In practice, the effects of different LDRs tend to cancel each other out to some extent (see e.g., the $\Tldr$ for Arcturus discussed in section 4 of T18).

\section{Summary}
\label{sec:summary}

We discussed the gravity effect on the LDR--$\Teff$ relations in the infrared $YJ$-band using 63 stars including dwarfs, giants, and supergiants.
Considering the line pairs selected in T18, some of the LDR--$\Teff$ relations of dwarfs are clearly shifted compared to the relations of giants and supergiants.
The pairs with line(s) affected by line blends tend to be sensitive to the surface gravity, but the gravity effect exists even without the blends.
We found that the difference between the ionization potentials of the elements involved in each line pair leads to different reactions of the line depths to $\logg$ and thus introduces the gravity effect.

To minimize the impact of the gravity effect and to increase the precision of the $\Teff$ derived with LDRs, we considered the relations calibrated separately for different luminosity classes.
We constructed new sets of LDR--$\Teff$ relations for dwarfs and supergiants using the lines selected for individual groups.
The residual around each relation ranges from 60 to 150\,K, and the final $\Tldr$ errors are around 20\,K.
Combined with the relations for giants in T18, which are well consistent with our measurements, it is possible to determine $\Teff$ for solar-metallicity dwarfs, giants, and supergiants precisely using the LDR method applied to $YJ$-band spectra. 
The $\Teff$ ranges that are covered by the available LDR--$\Teff$ relations in the $YJ$ bands are limited: 4500--6000\,K for dwarfs, 4000--5000\,K for giants, and 4000--6000\,K for supergiants.
More efforts are needed to extend the $\Teff$ ranges to cover and also to improve the relations with a larger number of calibrating stars.

\section*{Acknowledgements}

We are grateful to the staff of Koyama Astronomical Observatory for their support during our observation. 
This work has made use of the VALD database, operated at Uppsala University, the Institute of Astronomy RAS in Moscow, and the University of Vienna. 
This study has been financially supported by Grants-in-Aid (numbers 16684001, 20340042, 21840052, 26287028, and
18H01248) from the Japan Society for the Promotion of Science (JSPS) and by Supported Programs for the Strategic Research Foundation at Private Universities (S0801061 and S1411028) from the Ministry of Education, Culture, Sports, Science and Technology (MEXT) of Japan. 
KF is supported by a JSPS Grant-in-Aid for Research Activity Start-up (No. 16H07323). 
NK is supported by JSPS-DST under the Japan-India Science Cooperative Programs in 2013--2015 and 2016--2018.
We thank the referee, Dr. Ricardo P. Schiavon, for the comments which helps us  improve this paper.

\section*{Supporting Information}

Additional Supporting Information may be found in the on-line version of this article:

\textbf{Table~1.} Stellar parameters of the dwarfs in our sample together with their observation dates.

\textbf{Table~2.} Stellar parameters of the giants in our sample together with their observation dates.

\textbf{Table~3.} Stellar parameters of the supergiants in our sample together with their observation dates.

\textbf{Table~4.} The offsets, $\Delta \log{\mathrm{LDR}}$, between the LDR--$\Teff$ relations with the T18 line pairs of two luminosity classes measured at $\Teff=5000$\,K for the dwarf/supergiant and dwarf/giant pairs or at 4500\,K for the giant/supergiant pair. 

\textbf{Table~6.} The line pairs and the LDR--$\Teff$ relations, newly selected in this work, for dwarfs.

\textbf{Table~7.} The line pairs and the  LDR--$\Teff$ relations, newly selected in this work, for supergiants.

\textbf{LP\_all\_dwarf.pdf}, LDR--$\Teff$ relations of all the line pairs, supplementing Fig.~\ref{fig:LP_example_dwarf}, for dwarfs.

\textbf{LP\_all\_supergiant.pdf}, LDR--$\Teff$ relation plot of all the line pairs, supplementing Fig.~\ref{fig:LP_example_dwarf}, for supergiants.

\textbf{ir\_ldr-0.2.0.tar.gz}, The python package to measure line depths and LDRs and to calculate the temperatures, $\Tldr$.




\bibliographystyle{mnras}
\bibliography{refs.bib} 




\appendix


\bsp	
\label{lastpage}
\end{document}